\documentclass[fleqn,usenatbib]{mnras}

\usepackage{ae,aecompl}

\usepackage{graphicx}	
\usepackage{amsmath}	
\usepackage{amssymb}	
\usepackage{subcaption}

\title[IOPF: Oligarchic Coagulation of Planetesimals?]{Inside-Out Planet Formation: VI. Oligarchic Coagulation of Planetesimals from a Pebble Ring?}

\author[Cai, Tan \& Portegies Zwart]{
Maxwell X. Cai,$^{1}$\thanks{E-mail: cai@strw.leidenuniv.nl (MXC)}
Jonathan C. Tan,$^{2,3}$, and Simon Portegies Zwart$^{1}$
\\
$^{1}$Leiden Observatory, Leiden University, PO Box 9513, 2300 RA, Leiden, The Netherlands\\
$^{2}$Department of Space, Earth and Environment, Chalmers University of Technology, Gothenburg, Sweden \\
$^{3}$Department of Astronomy, University of Virginia, Charlottesville, VA 22904, USA
}

\date{Accepted 2021 December 8. Received 2021 December 8; in original form 2021 April 7}

\pubyear{2021}

\begin{document}
\label{firstpage}
\pagerange{\pageref{firstpage}--\pageref{lastpage}}
\maketitle

\begin{abstract}
Inside-Out Planet Formation (IOPF) is a theory addressing the origin of Systems of Tightly-Packed Inner Planets (STIPs) via {\it in situ} formation and growth of the planets. It predicts that a pebble ring is established at the pressure maximum associated with the dead zone inner boundary (DZIB) with an inner disk magnetorotational instability (MRI)-active region. Using direct $N$-body simulations, we study the collisional evolution of planetesimals formed from such a pebble ring, in particular examining whether a single dominant planet emerges. We consider a variety of models, including some in which the planetesimals are continuing to grow via pebble accretion. We find that the planetesimal ring undergoes oligarchic evolution, and typically turns into 2 or 3 surviving oligarchs on nearly coplanar and circular orbits, independent of the explored initial conditions or form of pebble accretion. The most massive oligarchs typically consist of about $70\%$ of the total mass, with the building-up process typically finishing within $\sim 10^5$ years. However, a relatively massive secondary planet always remains with $\sim30-65\%$ of the mass of the primary. Such secondary planets have properties that are inconsistent with the observed properties of the innermost pairs of planets in STIPs. Thus, for IOPF to be a viable theory for STIP formation, it needs to be shown how oligarchic growth of a relatively massive secondary from the initial pebble ring can be avoided. We discuss some potential additional physical processes that should be included in the modeling and explored as next steps.

\end{abstract}



\section{Introduction}\label{sec:introduction}

Thousands of exoplanets have been discovered in the last decade. Many
of these are in multi-transiting systems with tightly-packed inner
planets (STIPs) \citep{2012ApJ...761...92F}, which consist of $\sim1$ to $30\:R_\oplus$ radii planets found within $\sim 1$~au of their solar-type stellar host.
For example, the
Kepler-11 system consists of 6 super-Earths packed within 0.5~au from
the host star \citep{2011Natur.470...53L}, which is completely
different from our Solar System's orbital architecture.

Broadly speaking, the possible formation mechanisms of STIPs fall into
two categories: migration and \emph{in-situ} formation. For example, 
\cite{2012ARA&A..50..211K} suggest that STIPs
may be a consequence of disk-induced planet \emph{migration}. However, the
inward migration scenario predicts that planets become trapped in
low-order mean-motion resonances, which is inconsistent with
observations of STIPs \citep{2014prpl.conf..667B,2014ApJ...790..146F}.
Alternatively, {\it in situ} formation scenarios
have been proposed \citep{2013ApJ...775...53H,2013MNRAS.431.3444C,2014ApJ...780...53C} 
in which the planets form and grow in the inner disk. In particular, \cite{2014ApJ...780...53C} proposed the
\emph{Inside-Out Planet Formation} (IOPF) theory in which the planets
are formed sequentially from a pebble ring at the dead zone inner
boundary (DZIB) with an inner MRI-active zone. The pebble ring
coagulates and eventually builds up a single, dominant
planet. Subsequently, the DZIB retreats and a second pebble ring
forms, which in turn becomes the second planet, and so on.

This sequential formation of planets in the inner disk in the IOPF
scenario relies on the transport of pebbles through their
gas-drag-induced inward migration
\citep[e.g.,][]{1977MNRAS.180...57W}. Once pebbles are trapped in a
ring at the DZIB pressure maximum, the subsequent evolution is
expected to be a collisional process. It is thus of theoretical
importance to study this detailed collisional process. However, due to
the very large number of pebbles and frequent close encounters, it is
computationally expensive to investigate numerically. Previous
relevant studies include a pioneering work of
\cite{1986ApJ...305..564L}, who investigated the formation of
terrestrial planets through the inelastic collisions of 200
lunar-sized equal-mass planetesimals in a 2D plane. Later, making use
of the HARP-2 special-purpose supercomputer,
\cite{1998Icar..131..171K} carried out direct $N$-body simulations of
the coagulation of up to $N=4,000$ equal-mass planetesimals for
$10^4$~yr, and found that oligarchs form with a typical separation of
$\sim 10$ Hill radii. A few years later, the same authors carried out
larger simulations with $N=10^4$ equal-mass planetesimals for $4\times
10^5$~yr \citep{2002ApJ...581..666K}. Nevertheless, direct $N$-body
simulations remain highly expensive, even with modern hardware, and
multiple authors have developed alternative algorithms to mitigate the
computational costs. For example, \cite{2000Icar..143...45R} use the
parallel tree code \texttt{pkdgrav} to simulate $N=10^6$ particles for
1,000~yr on supercomputers; \cite{2007A&A...470..733S} modelled the
detailed outcome of two-body collisions and fragmentation of particles
using SPH simulations; based on a modified version of
\texttt{NBODY6++}, \cite{2014MNRAS.445.3620G} developed a hybrid
integration scheme that combines a Fokker-Planck approach with direct
$N$-body summation for handling large numbers of planetesimals.

In the IOPF series of papers, topics of planet-disk interaction
\citep{2016ApJ...816...19H}, pebble delivery and evolution
\citep{2018ApJ...857...20H}, and the inner-disk structure
\citep{2018ApJ...861..144M} have been considered. The scope of this
paper is to use direct $N$-body simulations to investigate the
build-up process of a planet from a ring of planetesimals that has
already formed from the pebble ring at the DZIB.

This paper is organised as follows: the modeling and initial conditions
are described in \S\ref{sec:modeling_ic}; results are
presented in \S\ref{sec:results}, followed by their discussion in
\S\ref{sec:discussion}; finally, conclusions are summarized
in \S\ref{sec:conclusions}.

\section{Initial Conditions and Modeling}
\label{sec:modeling_ic}

\subsection{Initial Conditions}

\cite{2014ApJ...780...53C} discuss that a typical location for the
DZIB is at $\sim 0.1$~AU for the first planet forming via IOPF around
a solar-type star. Then a pebble ring is expected to build up at the
pressure maximum at this location. Due to the low mass scale of
individual pebbles, direct modeling of the pebble ring requires
calculating the mutual gravity and the collisions of more than
$\sim10^{20}$ particles, which is far beyond the capability of direct
$N$-body simulations. Planetesimals may form from pebbles, e.g., via
the streaming instability \citep[][]{2005ApJ...620..459Y,
  2007Natur.448.1022J, 2008ApJ...687.1432C}. Therefore, for our
initial conditions we assume this process has already occurred and we
model the ring as a collection of planetesimals, i.e., we start with a planetesimal ring. We adopt an
initial total mass of the ring of $1\:M_{\oplus}$, i.e., enough material to be able to form a close-in planet comparable in mass to those observed in STIPs. For reasons of computational efficiency we carry out simulations involving an initial number of planetesimals equal to 100, i.e., approximately lunar-mass objects on average, although our power law models, described below, extend down to $0.001\:M_\oplus$. Thus a caveat of our work is that we start with relatively massive planetesimals and the range of mass scales followed in the simulations is limited to about a factor of 1000.

To investigate the dependence of the evolution on the initial mass
distribution of the planetesimals, we consider two cases: (1) an equal
mass distribution where each planetesimal has $0.01\:M_{\oplus}$; (2)
a power-law mass distribution of the following form:
\begin{equation}
	\frac{dN}{dM} \propto M^{k},
	\label{eq:imf}
\end{equation}
where we adopt a value of $k=-2$. 
This value means that each decade in mass of the populations contains,
on average, the same fraction of the total mass. The minimum mass of
the distribution is set to $0.001\:M_{\oplus}$, while there is no
maximum mass limit imposed. However, in order to insure that the total mass of $1\:M_\oplus$ is achieved with 100 objects, a scaling factor is applied to each sampling of the mass function. 
Given the chaotic nature of
$N$-body systems, we launch 20 realizations of each model with
different random seeds.

The mean density of planetesimals is assumed to be $2\:\textrm{g\:cm}^{-3}$,
which is a typical value for the main-belt asteroids
\citep{2012P&SS...73...98C}. We carried out test simulations of
planetesimals with a density of $4\:\textrm{g\:cm}^{-3}$, but we do not see
a significant differences compared to the results with
$2\:\textrm{g\:cm}^{-3}$.  Thus, the results in this paper are based on the
the fiducial choice of $2\:\textrm{g\:cm}^{-3}$ planetesimal density.

In the presence of gas, the orbital eccentricities and inclinations of
particles are expected to be kept small, but turbulence and/or
collisional processes could still lead to excitation of these
properties \citep[e.g.,][]{2016ApJ...822...54D}. For this reason, we
sample the initial eccentricities and inclinations from a Rayleigh
distribution, with an RMS value of $\langle e^2 \rangle^{1/2} =
2\langle i^2 \rangle^{1/2}=0.02$
\citep{1993prpl.conf.1061L,2002ApJ...581..666K}. The width of the
pebble ring $W \equiv a_{\rm out} - a_{\rm in}$ is another loosely
constrained parameter, although in the IOPF model it is expected to be limited by the
width of the DZIB region. In our simulations, we assume three
different values of $W=[0,0.01,0.02]$~au, i.e., from the limit of a very thin ring to one that has a width up to 20\% of its radius. We will show later, that our
results are insensitive to this choice of $W$.

\subsection{Treatment of Collisions}

For simplicity, we assume that planetesimal collisions are perfectly
inelastic and lead to formation of a new, spherical planetesimal that
has conserved the total linear momentum, mass and volume (but not the
energy), of its precursors.
This assumption is valid since collisional
fragmentation is not a barrier for planet formation in close-in orbits
\citep{2017AJ....154..175W}, such as those we use in the
simulation.

\subsection{Pebble Accretion}

While we do not directly model the process in which planetesimals are
initially formed from pebbles, for some of our simulations we do take
into account the effects of planetesimal growth by pebble
accretion. Due to the slightly sub-Keplerian orbital velocity of the
gas where the gas pressure is declining with radius, pebbles feel a
headwind drag and drift inwards from the outer regions of the disk
\citep{1977MNRAS.180...57W}. In the IOPF model the pebbles are assumed
to be able to reach the DZIB and the outer protoplanetary disk serves
as a reservoir of the pebbles.

If the pebble layer at the DZIB is \emph{thick} compared to the radius
of the planetesimal, then it is reasonable to assume that pebbles
accrete at arbitrary latitudes and from arbitrary directions onto the
planetesimal surface. Alternatively, if the pebble layer is
\emph{thin}, accretion takes place concentrated around the
heliocentric plane. Due to gravitational focusing, more massive
planetesimals will attract pebbles from larger initial orbital impact
parameters. Therefore, the mass-dependent accretion rate, which can be
expressed via a power law index $\beta$ in
Eq.~\ref{eq:pebble_accretion_power_index}, depends on both the
thickness of the pebble layer relative to planetesimal size and the
effect of gravitational focusing:
\begin{equation}
	\dot{M}_i \propto M^{\beta}_{i}
	\label{eq:pebble_accretion_power_index},
\end{equation}
where $M_i$ is the mass of the $i$-th planetesimal. Various values of
$\beta$ have been adopted in previous studies of pebble
accretion. For example, \cite{2010A&A...520A..43O} assumed $\beta =
2/3$, while \cite{2018MNRAS.479.5136P} used $\beta = 1$. Here, since
we find that the pebble layer is expected to be thick compared to the
size of the planetesimals, we will consider a case with $\beta = 4/3$
(see Appendix~\ref{appendix:thick_disk}). However, we will also
consider cases with no pebble accretion and an intermediate case with
$\beta=2/3$ (see Table~\ref{tab:ic} for a summary of
initial conditions).

For the models with pebble accretion, we assume a total pebble supply
rate to the DZIB region of $\dot{M}_{\rm p,acc}= 10^{-11}
M_{\odot}\:{\rm yr}^{-1}$, i.e., 1\% of the gas accretion rate
\citep[see also][]{2018ApJ...857...20H}. We then assume this is divided among the
planetesimals in proportion to the estimated individual accretion
rates that would be achieved in isolation, as described above, i.e.,
$\dot{M}_{\rm p,acc}= \sum_{1}^n(\dot{M}_i) = 10^{-11}
M_{\odot}\:{\rm yr}^{-1}$.

\subsection{Simulation Code and Duration}

The simulations are carried out with \texttt{ABIE}\footnote{\url{https://github.com/MovingPlanetsAround/ABIE}} (Cai et al. in prep), a new GPU-accelerated direct $N$-body
code. The integrator adopted in these simulations is a 15th-order
Gauss-Radau algorithm \citep{1985dcto.proc..185E} with an adaptive
timestep scheme. The algorithm is particularly optimized for close
encounters. Taking into account the dissipative energy during the
collision process, the relative energy error is of the order of
$dE/E_{0} \sim 10^{-7}$ by the end of the simulations\footnote{Note,
  however, that this energy conservation check is no longer useful for
  models with pebble accretion.}, which is sufficient to warrant a valid numerical solution to the gravitational $N$-body problem \citep{2015ComAC...2....2B}.

All simulations are carried out for 1~Myr, which is approximately $3
\times 10^7$ orbits at $a \sim 0.1$~AU. During this time period, the
models with pebble accretion are supplied with $10^{-5}\:M_\odot =
3.3\:M_\oplus$.

An overview of the simulated models is presented in
Table~\ref{tab:ic}. Each model has 20 realizations with different
random seeds. All simulations were carried out automatically using
Simulation Monitor \texttt{SiMon}\footnote{\url{https://github.com/maxwelltsai/SiMon}} \citep{2017PASP..129i4503Q}.

\begin{table}
	\centering
	\begin{tabular}{l|r|r|r|r}
		\hline
		\hline
		Model                & $a_{\rm in}$~[au] & $W$ [au] & $\beta$\\
		\hline
		\texttt{EMS\_0.00\_NPA} & 0.1 & 0        & --        \\
		\texttt{EMS\_0.01\_NPA} & 0.1 & 0.01     & --        \\
		\texttt{EMS\_0.02\_NPA} & 0.1 & 0.02     & --        \\
		\hline 
		\texttt{PWR\_0.00\_NPA}& 0.1 & 0        & --       \\
		\texttt{PWR\_0.01\_NPA}& 0.1 & 0.01     & --       \\
		\texttt{PWR\_0.02\_NPA} & 0.1 & 0.02     & --       \\
		\hline
		
		\texttt{EMS\_0.00\_PA\_2\_3} & 0.1 & 0        & $2/3$        \\
		\texttt{EMS\_0.01\_PA\_2\_3} & 0.1 & 0.01     & $2/3$        \\
		\texttt{EMS\_0.02\_PA\_2\_3} & 0.1 & 0.02     & $2/3$        \\
		\hline 
		\texttt{PWR\_0.00\_PA\_2\_3} & 0.1 & 0        & $2/3$        \\
		\texttt{PWR\_0.01\_PA\_2\_3} & 0.1 & 0.01     & $2/3$        \\
		\texttt{PWR\_0.02\_PA\_2\_3} & 0.1 & 0.02     & $2/3$        \\
		\hline 	
		
		\texttt{EMS\_0.00\_PA\_4\_3} & 0.1 & 0        & $4/3$        \\
		\texttt{EMS\_0.01\_PA\_4\_3} & 0.1 & 0.01     & $4/3$        \\
		\texttt{EMS\_0.02\_PA\_4\_3} & 0.1 & 0.02     & $4/3$        \\
		\hline 
		\texttt{PWR\_0.00\_PA\_4\_3} & 0.1 & 0        & $4/3$        \\
		\texttt{PWR\_0.01\_PA\_4\_3} & 0.1 & 0.01     & $4/3$        \\
		\texttt{PWR\_0.02\_PA\_4\_3} & 0.1 & 0.02     & $4/3$        \\
		\hline
		\hline
	\end{tabular}
	\caption{
Initial conditions of the models adopted in this study. Models with
the name \texttt{EMS*} are the equal mass systems, and with the name
\texttt{PWR*} are the non-equal-mass systems with a power-law initial
mass function ($\k=-2$).
$\beta$ is the
mass-dependent power law index for gravitational-focusing-enhanced pebble
accretion (when it is blank, there is no pebble accretion). Each model is simulated with 20 realizations, resulting in a
total of 360 simulations. }
	\label{tab:ic}
\end{table}

We note that our $N$-body simulations do not include any forces on the planetesimals due to the effects of gas, i.e., gas drag or gravitational forces. We discuss the potential effects of these limitations in \S\ref{sec:discussion}.

\section{Results}
\label{sec:results}

\subsection{Mass Distribution of Surviving Planets}

\begin{figure*}
    \centering
    \includegraphics[scale=0.5]{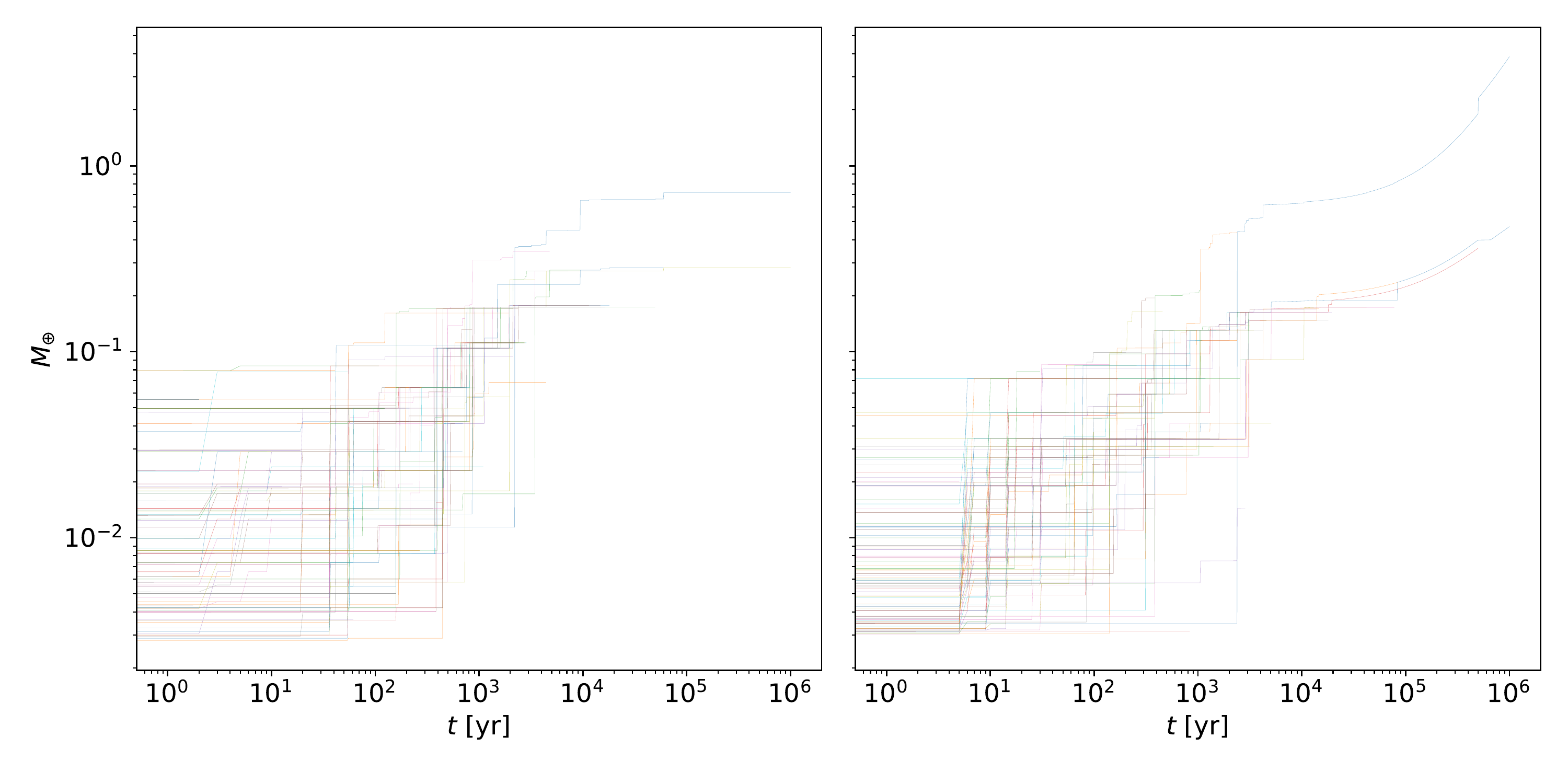}
    \caption{Planetesimal mass growth versus time in example simulations. Each colored line represents a planetesimal. If two planetesimals merge, then the more massive one retains its identity (i.e., line color). {\it (a) Left:} an example simulation without pebble accretion. The total mass of planetesimals is $1 M_{\oplus}$, distributed to 100 planetesmals following the power-law initial mass distribution described in Eq.~\ref{eq:imf}. {\it (b) Right:} same initial conditions parameters as (a), but with pebble accretion following Eq.~\ref{eq:pebble_accretion_power_index} with $\beta = 4/3$. The total pebble accretion rate is  $\dot{M}_{\rm p, acc} = 10^{-11} M_{\odot} \: {\rm yr}^{-1}$. }
    \label{fig:accretion_history}
\end{figure*}

\begin{figure}
	\centering
	\includegraphics[scale=0.55]{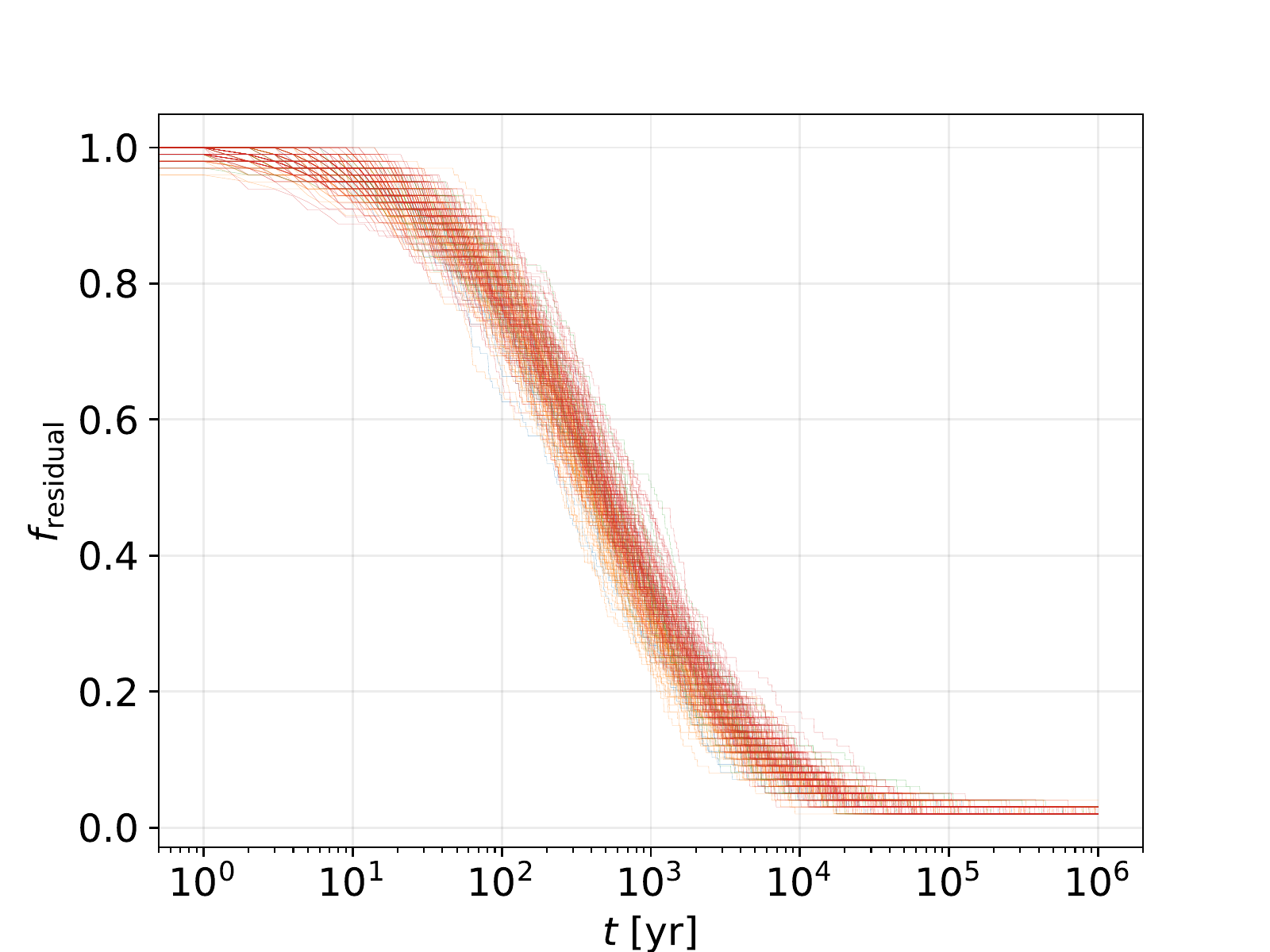}
	\caption{
	Time evolution of the fraction of residual particles, i.e., ratio of number of surviving planetesimals to the initial number. The various different simulation types explored in this paper show similar behavior, i.e., with just a few planets remaining after $\sim10^5\:$yr.
	}
	\label{fig:f_bound}
\end{figure}

We find that, regardless of the initial conditions, most simulations (84.5\%) end up with only two surviving planets. Almost all the other systems (15.3\%) have three surviving planets, with just one system (0.2\%) having four planets left at the end of the simulation. Figure~\ref{fig:accretion_history} shows the time evolution of planetesimal mass growth and merging in two example simulations. In the first example, which is a case without pebble accretion, the system ends up with two surviving planets, with this state achieved after about 60,000~yr. In the second example, with pebble accretion, the system also ends up with two surviving oligarchs, but this state is only reached after about $5\times10^5\:$yr (although the three-planet configuration is reached within $10^5\:$yr). Figure~\ref{fig:f_bound} shows the time evolution of the fraction of residual planetesimals, i.e., the ratio of the number of those surviving to the initial number, for the ensemble of simulations investigated in this paper. The same general trend is seen, i.e., that an oligarchic state with just a few surviving planets is typically achieved within about $10^5\:$yr.

To understand whether or not the majority of the mass is concentrated into a dominant protoplanet or is more evenly distributed over multiple planets, for the conditions at the final time of $10^6\:$yr, we consider the ratio of the mass of the most massive surviving planet (hereafter: the primary), $M_{\rm pri}$, to the total mass of the planets, $M_{\rm total}$. We also consider the ratio of $M_{\rm pri}$ to that of the second most massive planet (hereafter: the secondary), $M_{\rm sec}$.  The minimum, maximum, mean and median values of these ratios are listed in Table~\ref{tab:mr}.

\begin{table*}
\caption{Ratios of primary mass to secondary mass ($M_{\rm pri} / M_{\rm sec}$) and primary mass to total mass ($M_{\rm pri} / M_{\rm total}$).}
\begin{tabular}{l|rrrr|rrrr}

\hline
\hline
{} & \multicolumn{4}{c}{$M_{\rm pri} / M_{\rm sec}$} & \multicolumn{4}{c}{$M_{\rm pri} / M_{\rm total}$} \\
\cline{2-5} \cline{6-9}
{Model} &   min &   max &  mean & median &   min &   max &  mean & median \\

\hline
EMS\_0\_NPA         &  1.09 &  5.10 &  1.69 &   1.32 &  0.48 &  0.84 &  0.59 &   0.57 \\
EMS\_0.01\_NPA      &  1.06 &  6.59 &  2.32 &   2.01 &  0.51 &  0.87 &  0.65 &   0.61 \\
EMS\_0.02\_NPA      &  1.10 &  7.28 &  2.57 &   2.35 &  0.48 &  0.88 &  0.66 &   0.64 \\
{\bf (Average)} &  1.08 &  6.32 &  2.19 &   1.89 &  0.49 &  0.86 &  0.63 &   0.61 \\
\hline
PWR\_0\_NPA          &  1.07 &  4.38 &  2.08 &   1.47 &  0.51 &  0.81 &  0.63 &   0.58 \\
PWR\_0.01\_NPA       &  1.02 &  2.72 &  1.71 &   1.69 &  0.50 &  0.73 &  0.61 &   0.63 \\
PWR\_0.02\_NPA       &  1.03 &  3.73 &  1.96 &   1.85 &  0.49 &  0.79 &  0.62 &   0.59 \\
{\bf (Average)} &  1.04 &  3.61 &  1.92 &   1.67 &  0.50 &  0.78 &  0.62 &   0.60 \\
\hline
EMS\_0\_PA\_2\_3    &  1.03 &  4.80 &  1.52 &   1.33 &  0.47 &  0.83 &  0.57 &   0.54 \\
EMS\_0.01\_PA\_2\_3 &  1.06 &  3.35 &  1.53 &   1.36 &  0.44 &  0.77 &  0.58 &   0.57 \\
EMS\_0.02\_PA\_2\_3 &  1.03 &  2.73 &  1.44 &   1.26 &  0.44 &  0.73 &  0.57 &   0.54 \\
{\bf (Average)} &  1.04 &  3.62 &  1.50 &   1.31 &  0.45 &  0.78 &  0.57 &   0.55 \\
\hline
PWR\_0\_PA\_2\_3    &  1.00 &  5.19 &  1.99 &   1.52 &  0.50 &  0.84 &  0.62 &   0.60 \\
PWR\_0.01\_PA\_2\_3 &  1.10 &  4.77 &  1.68 &   1.41 &  0.52 &  0.83 &  0.60 &   0.59 \\
PWR\_0.02\_PA\_2\_3 &  1.04 &  2.10 &  1.34 &   1.20 &  0.51 &  0.68 &  0.56 &   0.54 \\
{\bf (Average)} &  1.05 &  4.02 &  1.67 &   1.38 &  0.51 &  0.78 &  0.59 &   0.58 \\
\hline
EMS\_0\_PA\_4\_3    &  1.03 &  13.48 &  4.52 &   3.55 &  0.51 &  0.92 &  0.75 &   0.77 \\
EMS\_0.01\_PA\_4\_3 &  1.11 &   8.02 &  2.68 &   1.95 &  0.53 &  0.89 &  0.69 &   0.66 \\
EMS\_0.02\_PA\_4\_3 &  1.02 &   5.77 &  2.87 &   2.37 &  0.51 &  0.85 &  0.70 &   0.70 \\
{\bf (Average)} &  1.05 &   9.09 &  3.36 &   2.62 &  0.52 &  0.89 &  0.71 &   0.71 \\
\hline
PWR\_0\_PA\_4\_3    &  1.11 &  11.27 &  3.42 &   1.94 &  0.53 &  0.92 &  0.71 &   0.66 \\
PWR\_0.01\_PA\_4\_3 &  1.04 &   8.20 &  3.45 &   2.71 &  0.51 &  0.89 &  0.72 &   0.71 \\
PWR\_0.02\_PA\_4\_3 &  1.35 &  10.09 &  3.40 &   2.42 &  0.57 &  0.90 &  0.71 &   0.71 \\
{\bf (Average)} &  1.17 &   9.85 &  3.42 &   2.36 &  0.54 &  0.90 &  0.71 &   0.69 \\
\hline
\end{tabular}
\label{tab:mr}
\end{table*}

The mean values of $M_{\rm pri}/M_{\rm total}$ range from 0.56 to 0.75
depending on the initial conditions and whether or not pebble
accretion is included. Within each set of simulations for a particular
case there is a greater range, with a minimum value of 0.44 and
maximum value of 0.92 seen as the result of individual simulation
runs. Thus we can see that usually most of the total mass is
concentrated into a single most massive planet.

The extent to which the primary dominates its local environment is
best judged by considering the values of $M_{\rm pri}/M_{\rm sec}$. Here we see more clearly that there are differences depending on initial conditions of the simulation. In particular, if pebble accretion is included then
$M_{\rm pri}/M_{\rm sec}$ tends to be greater, and this is driven by the adoption of mass-dependent pebble accretion rates, i.e., more massive
planetesimals accrete at higher rates. Thus $M_{\rm pri}/M_{\rm sec}$ tends to be higher for the $\beta=4/3$ case than the $\beta=2/3$ case. On the
other hand, whether the population starts with equal masses or a power
law distribution does not have a major systematic influence on the
final values of $M_{\rm pri}/M_{\rm sec}$. There is also no large systematic effect on $M_{\rm pri}/M_{\rm sec}$ of the initial width of the planetesimal distribution in orbital radius.

\begin{figure*}
	\centering
	\includegraphics[scale=0.55]{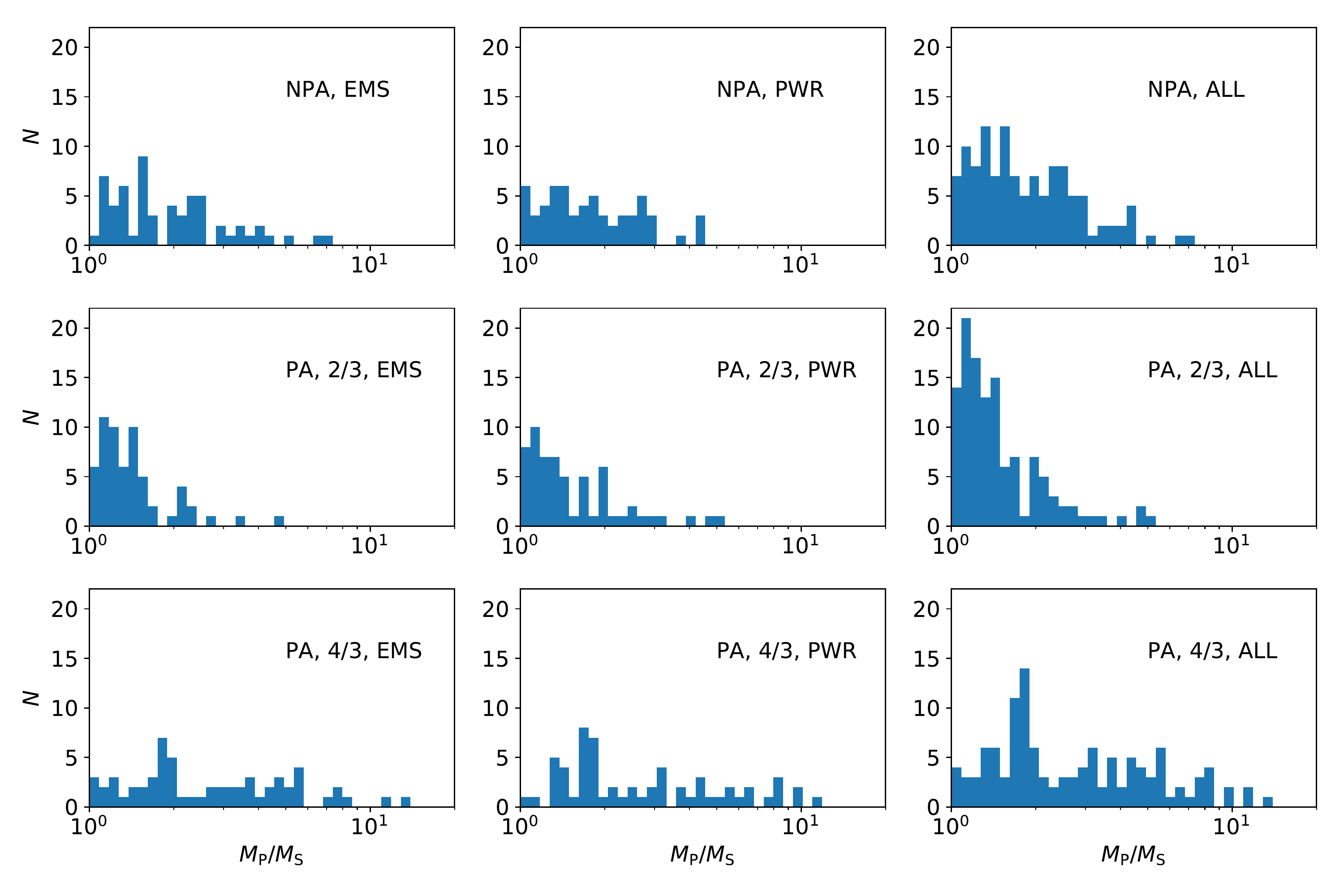}
	\caption{
Distribution of ratios of primary mass to secondary mass,
$M_{\rm pri}/M_{\rm sec}$, binned in equal intervals of log$_{10}(M_{\rm pri}/M_{\rm sec}$).
The results of the \texttt{EMS} model, the \texttt{PWR} model and the
combination of \texttt{EMS+PWR} models are plotted in the first,
second and third column, respectively. The results for the model
without pebble accretion (NPA), with pebble accretion with
$\beta=2/3$, and with pebble accretion with $\beta=4/3$ are shown in
the first, second and third rows, respectively. }
	\label{fig:mass_dist_hist_lin_x}
\end{figure*}

We visualize the distributions of $M_{\rm pri}/M_{\rm sec}$ in
Fig.~\ref{fig:mass_dist_hist_lin_x}, where we have condensed all the
different initial conditions listed in Table~\ref{tab:ic} to six different models, i.e., no pebble accretion (NPA, top row), pebble accretion (PA) with
$\beta=2/3$ (middle row) and PA with $\beta=4/3$ (bottom row), for the
equal mass (EMS, left column) and power law (PWR, middle column)
initial mass distributions. Since EMS and PWR have quite similar
outcomes, we also show the summed distributions (EMS+PWR) of all these
cases in the right column of Fig.~\ref{fig:mass_dist_hist_lin_x}. Note
that in these panels we are binning in equal intervals of
log$_{10}(M_{\rm pri}/M_{\rm sec})$. 

The distribution in values of $M_{\rm pri}/M_{\rm sec}$ of the three cases NPA, PA:$\beta=2/3$ and PA:$\beta=4/3$ are clearly distinct, with the
latter extending to highest values and having the largest
dispersions. In these aspects, next in the sequence are the NPA
distributions, followed by the PA:$\beta=2/3$ distributions, which
tend to have lower values and smaller dispersions. Thus the index of
pebble accretion, i.e., whether $\beta>1$ or $<1$ plays a role in the
extent to which the primary dominates with respect to the secondary.

\subsection{Orbital Properties of Surviving Planets}

\begin{figure*}
	\centering
	\includegraphics[scale=0.7]{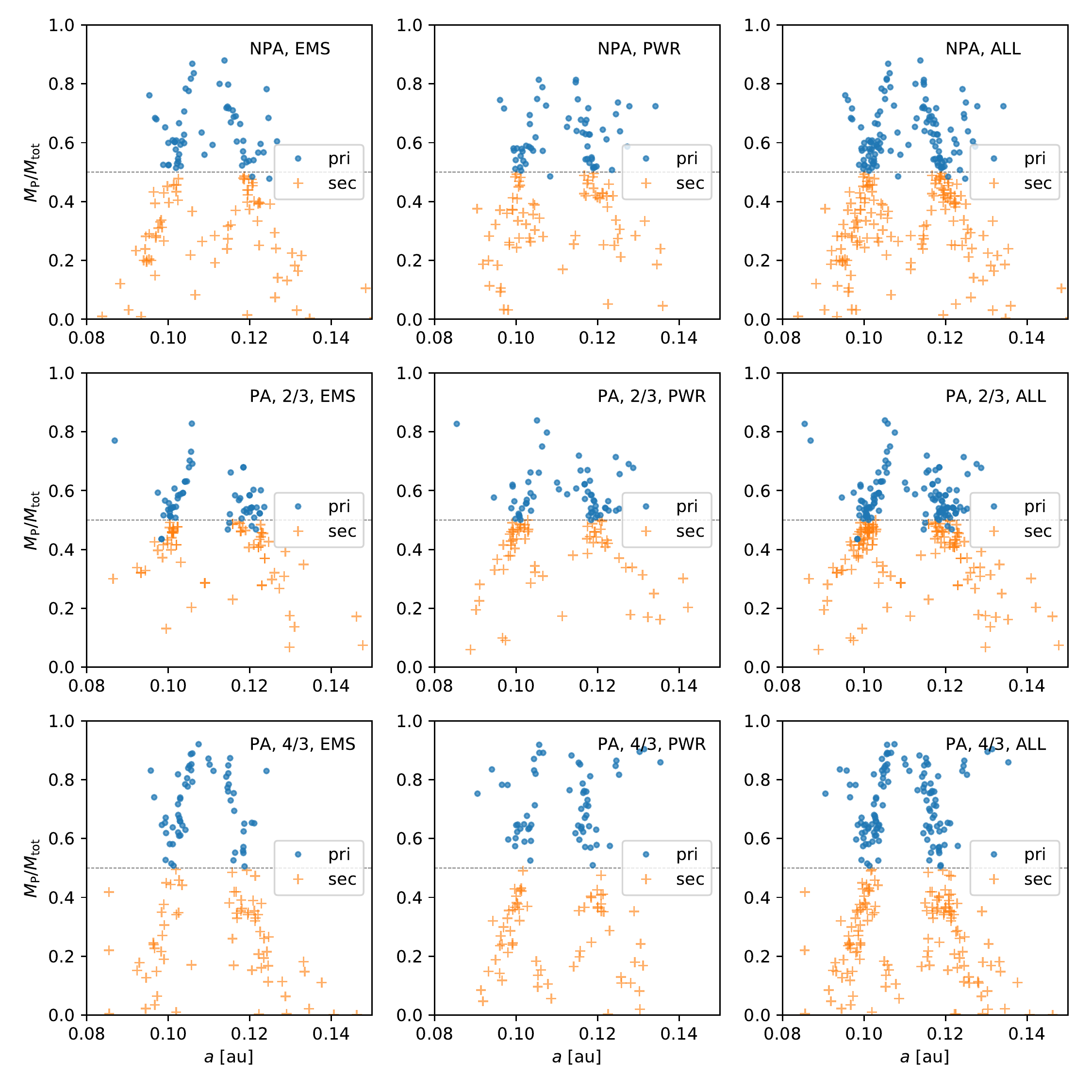}
	\caption{
The final masses (relative to total) versus semi-major axes of all
surviving planets. Primaries are colored blue; secondaries are
colored orange. Most primaries are more massive than half of the total mass (indicated by the dashed line), but there is 
a small fraction of primaries having less than half of the total mass, i.e., some of those that are in 3-planet systems.  
The ordering of the subplot grid is the same as
Fig.~\ref{fig:mass_dist_hist_lin_x}. Primaries have a narrower distribution in semi-major axis than the secondaries.
}
	\label{fig:mass_dist_semi}
\end{figure*}

Figure~\ref{fig:mass_dist_semi} shows the masses (normalized to
$M_{\rm total}$) versus orbital semi-major axis of the final
planets, with separate panels for the nine cases, following the
format of Fig.~\ref{fig:mass_dist_hist_lin_x}. Note the initial
distributions all start at $a_{\rm inner}=0.1$~au and extend out to
0.12~au in the widest case. Thus, as a result of orbital interactions,
there is a modest spreading in the range of semi-major axes
(accompanied by an increase in orbital eccentricities, see
below). There are apparent clusterings at particular values of $a$,
e.g., around 0.1~au and 0.12~au, which we discuss below in terms of the implied period ratios. Overall, one of the main features to be noted is that primaries have a narrower distribution in semi-major axis than the secondaries.

\begin{figure*}
	\centering
	\includegraphics[scale=0.7]{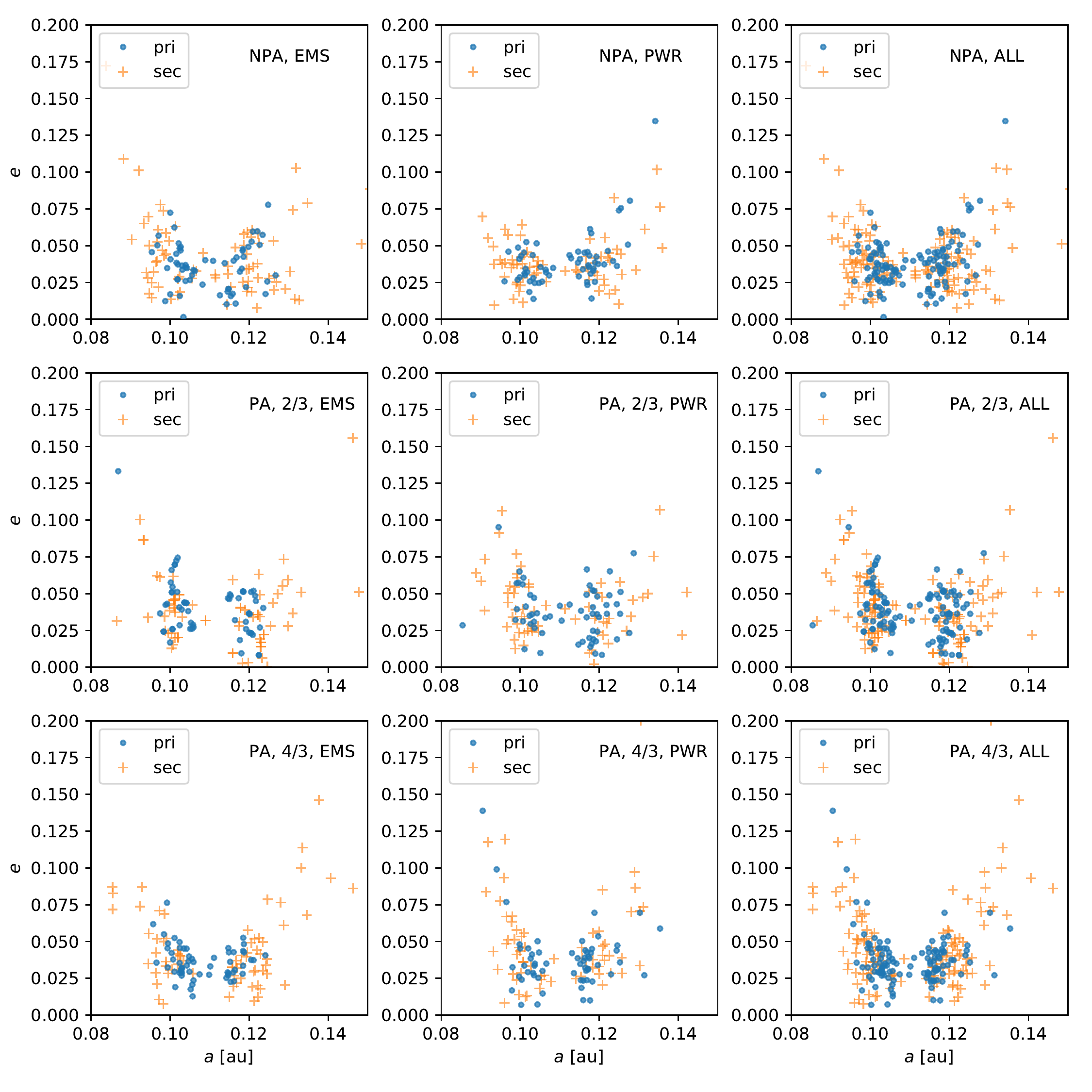}
	\caption{
Scatter plots of orbital eccentricity, $e$, versus semi-major axis,
$a$, for the surviving protoplanets. The primaries, which are more massive 
than the secondaries, tend to have smaller orbital eccentricities. 
The ordering of the subplot grid
is the same as Fig.~\ref{fig:mass_dist_hist_lin_x}.}
	\label{fig:a_e_dist}
\end{figure*}

Figure~\ref{fig:a_e_dist} shows orbital eccentricity, $e$, versus
semi-major axis of the final planets, again with the same
nine-panel format as before. Modest degrees of eccentricity excitation
are seen, i.e., with typical values of $\sim0.03$, but extending up to about 0.15. As expected, those planets that are further in semi-major axis from the range of the initial conditions, i.e., $a=0.1$ to 0.12, have larger eccentricities. Thus, the primaries tend to have smaller eccentricities (and also inclinations, not shown), compared to the less massive planets. We note that there is no planetary ejection in any simulation; rather, the surviving oligarchs are generally settled on co-planar orbits. 

\begin{figure*}
    \centering
    \includegraphics[scale=0.38]{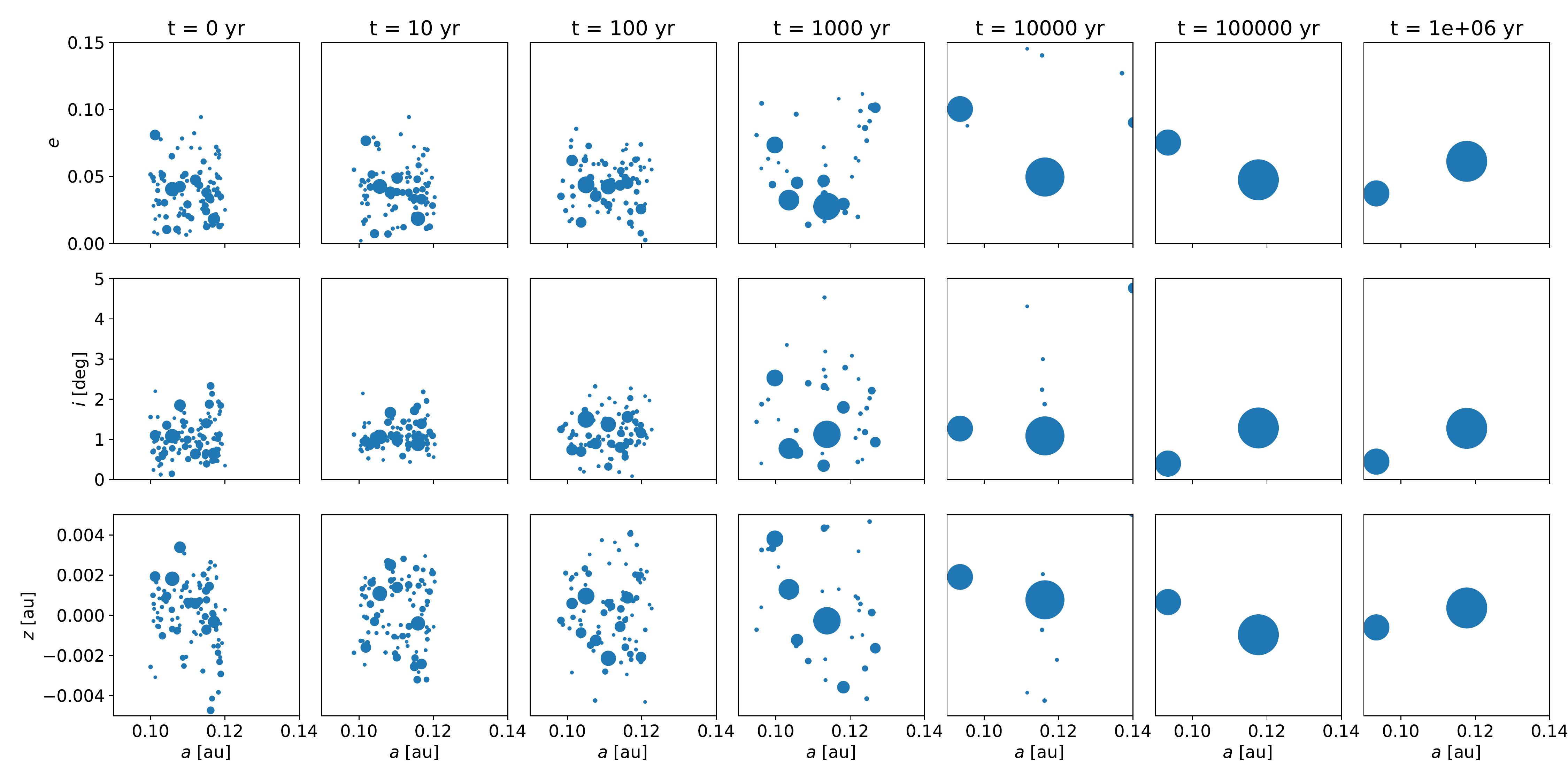}
    \caption(a) Top row: Seven snapshots at various times, as labelled, of the dynamical evolution and oligarchic coagulation of planetesimals at the DZIB in an example simulation, shown in eccentricity ($e$) versus semi-major axis ($a$) space. The size of the circles indicates the mass of the planetesimals. In this simulation the initial masses of planetesimals are sampled from a power-law with index of $k = -2$ and there is no pebble accretion included. 
    (b) Middle row: As (a), but now showing the inclination ($i$) versus semi-major axis ($a$) distributions of the planetesimals.
    (c) Bottom row: As (a), but now showing the vertical height ($z$) versus semi-major axis ($a$) distributions of the planetesimals.
    \label{fig:dyn_a_e}
\end{figure*}

To more fully illustrate the dynamical evolution that is occurring in the simulations, Figure~\ref{fig:dyn_a_e}a presents several snapshots of an example simulation (the same one shown in Fig. 1a, i.e., power law distribution of initial masses and no pebble accretion) in $a-e$ space, while planetesimal masses are indicated by the size of the circles. Figures~\ref{fig:dyn_a_e}b and c presents the equivalent results for the planetesimals in $a-i$ and $a-z$ space. During the entire process, the planetesimals are confined within a small island in the $a-e$ space. Here we can see that, as expected, it is the lower-mass planetesimals that can be excited to the largest eccentricities, although the range of these values is relatively modest.


When following the dynamical evolution in the simulations, i.e., of eccentricities, inclinations and vertical scale heights, we note that the average scale height of the planetesimals obtains values of $\sim 0.002\:$au (simple averaging by number) and $\sim 0.001\:$au (weighted averaging by masses) (see also Fig.~\ref{fig:dyn_a_e}c)\footnote{The averaged scale heights are changing during the first 100 kyr, after which most planetesimals are merged and the averaged scale heights remain approximately constant.}. These values are similar to the estimated vertical scaleheights of pebbles in the DZIB region of the pebble ring (see Appendix~\ref{appendix:thick_disk}), especially for pebbles with radii $\lesssim 1\:{\rm cm}$. Furthermore, this estimate of pebble scaleheight due to stirring by turbulence ignores any additional thickening due to the stirring gravitational influence of the planetesimals themselves. The fact that lower-mass planetesimals tend to have larger scaleheights that may become larger than the pebble scaleheight could lead to them having less efficient pebble accretion rates and tend to boost runaway growth of a dominant protoplanet. However, given the uncertainties in pebble properties at the DZIB, we have not attempted to include this effect in our modelling.


\begin{figure*}
	\centering
	\includegraphics[scale=0.7]{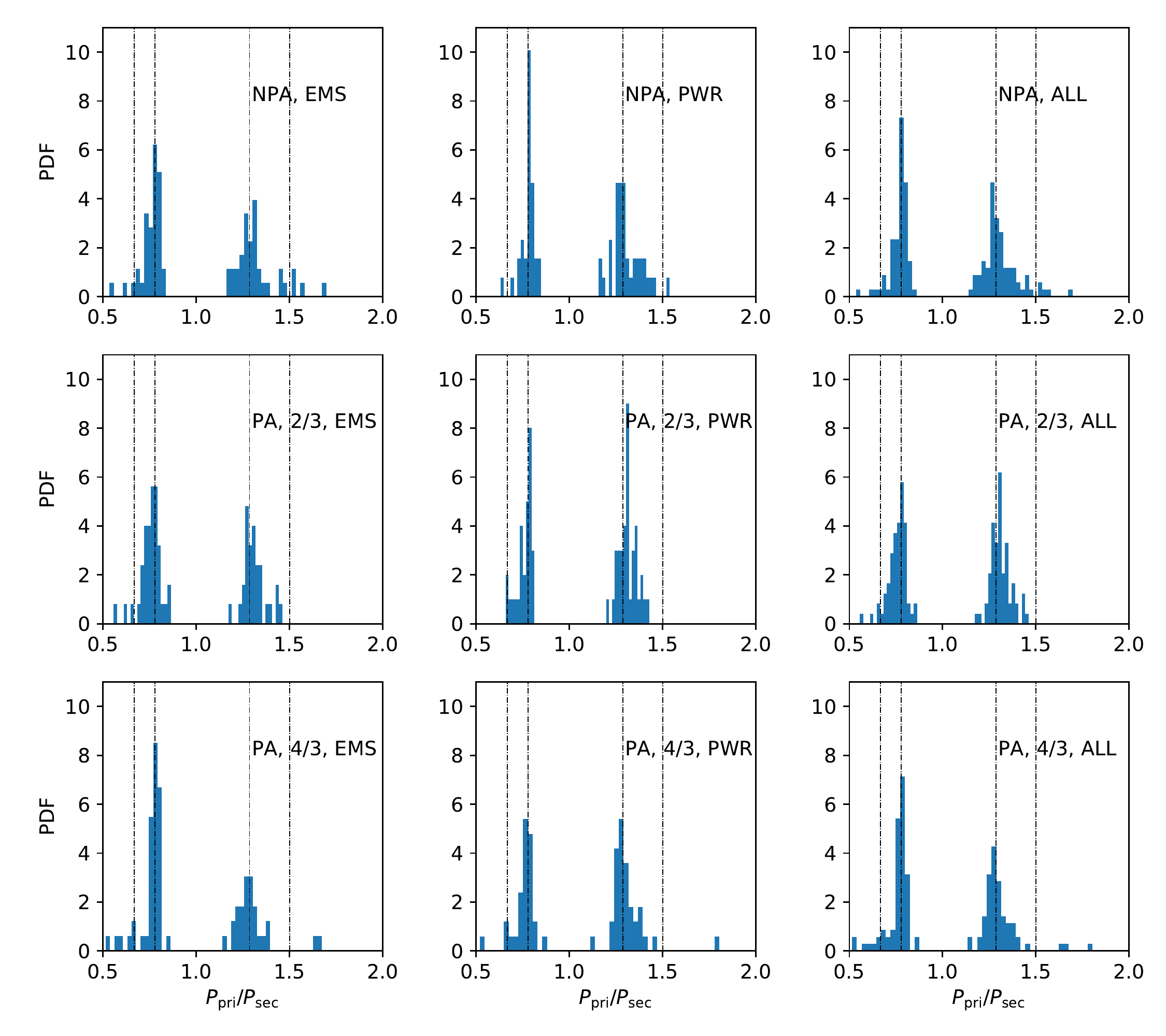}
	\caption{The probability distribution function (PDF) of the orbital period ratios between the primary and the secondary ($P_{\rm pri}/P_{\rm sec}$). The ordering of the subplot grid is the same as Fig.~\ref{fig:mass_dist_hist_lin_x}. In each panel, the four dashed vertical lines correspond to the mean motion resonance ratios of $2/3$, $7/9$, $9/7$, and $3/2$.}
	\label{fig:period_histogram}
\end{figure*}

Figure~\ref{fig:period_histogram} shows histograms of the distributions of $P_{\rm pri}/P_{\rm sec}$. The resonances have peaks at 7/9 and 9/7, respectively, rather than in a lower order (e.g., 2/3, 3/2). This is likely due to the assumed initial conditions of our simulations, i.e., with the planetesimals starting from a fairly narrow range of semi-major axes and eccentricities. However, note that it is still only a minority of systems that end up trapped in the 7/9 or 9/7 resonances.

\begin{figure*}
	\centering
	\includegraphics[scale=0.55]{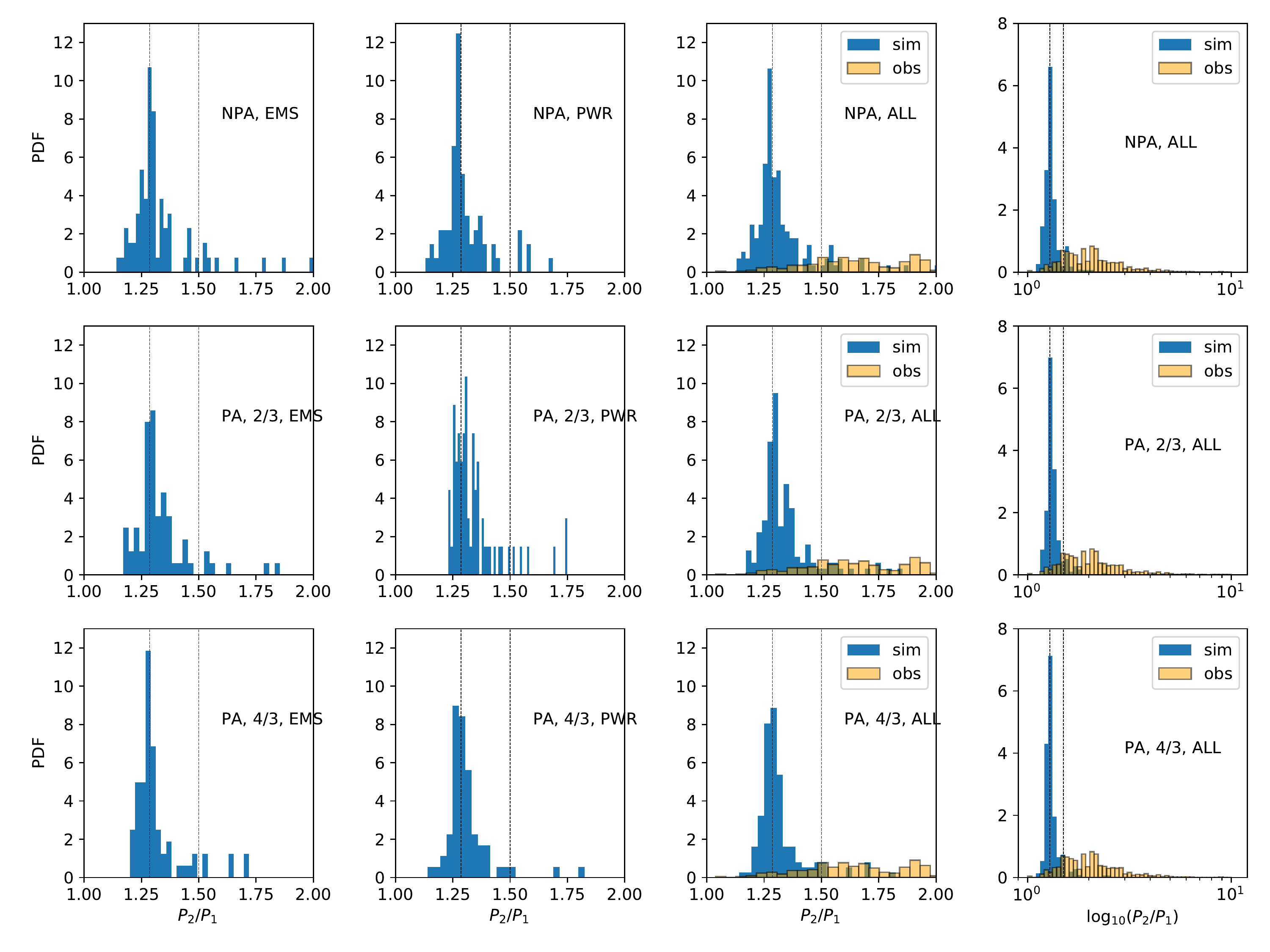}
	\caption{The probability distribution functions (PDFs) of the orbital period ratios between the innermost planet (1) and the second innermost planet (2) (denoted as $P_2/P_1$) for the various simulation sets, as labelled. The left-side 3 by 3 grid of panels follows the same layout as used in Figures~\ref{fig:mass_dist_hist_lin_x} to \ref{fig:period_histogram}, while the fourth column is the same as the third, but with an expanded horizontal range. Dotted vertical lines corresponding to the mean motion resonance ratios of $9/7$ and $3/2$ are shown in each panel. The observational data (see text) are plotted in the third and fourth columns for comparison. 
}
	\label{fig:period_histogram_p2_p1}
\end{figure*}

To allow easier comparison with observed systems, in Figure~\ref{fig:period_histogram_p2_p1} we plot the period ratios of the two innermost planets, $P_2/P_1$, i.e., where subscript 1 refers to the very innermost (``Vulcan'') planet and subscript 2 refers to the second closest planet from the star. Recall that since about 85\% of the surviving systems are two planet systems, these two innermost planets will generally be composed of the primary and secondary of each system. Here we see that our simulated planetary systems that formed from a narrow planetesimal ring end up with a narrow distribution of $P_2/P_1$, i.e., in the range from about 1.2 to about 2, with a peak at the 9/7 (=1.29) MMR occupied by a modest fraction of the systems.

Figure~\ref{fig:period_histogram_p2_p1} also shows the observational data for $P_2/P_1$ derived from the exoplanet archive\footnote{\url{http://exoplanet.eu/}}
for STIPs, i.e., following \cite{2014ApJ...780...53C} in selecting planets that are in systems with $N\geq2$ planets and excluding the few multiplanet systems with large ($\geq9\:R_\oplus$) planets.
As of 25 March 2021, there are 773 systems with two or more such STIPs in this database, among which 536 systems have well determined planetary radii and orbital periods, so our observational ``STIPs sample'' is based on these inner planet pairs.
It is evident that the simulation data is inconsistent with the observed systems. In our simulations, the $P_2/P_1$ ratio has a narrow distribution that peaks at 9/7. On the other hand, the observed systems have a more widely spread distribution (especially see the rightmost column of Fig.~\ref{fig:period_histogram_p2_p1}, which shows an expanded range of $P_2/P_1$). Note, that while most of the observed systems are not in low-order mean motion resonances, there is a modest fraction, $\sim10\%$, that are (see \S\ref{sec:introduction}). The third column of Fig.~\ref{fig:period_histogram_p2_p1}, which zooms in the range $1<P_2/P_1<2$, shows that there is an enhancement of the observed systems in the 3/2 MMR, but very few in the 9/7 MMR.

The implication of this discrepancy in the simulated and observed distributions is that the observed innermost planet pairs do not have orbital properties that are consistent with them being formed together via oligarchic growth from a single, narrow planetesimal ring. The sense of the discrepancy is that the simulated planets have a much narrower distribution of period ratios. The implications of this result for IOPF are discussed below in \S\ref{sec:discussion}, however we note here that the standard IOPF scenario involves a single planet forming from each pebble ring, with the orbital properties of next innermost planet being set by the location of the second pebble ring, which is set by retreat of the dead zone inner boundary. Our results here constrain how the first planet in the IOPF scenario can have formed and argue against there being a stage where there was oligarchic growth from a population of self-gravitating planetesimals.

\begin{figure*}
	\centering
	\includegraphics[scale=0.7]{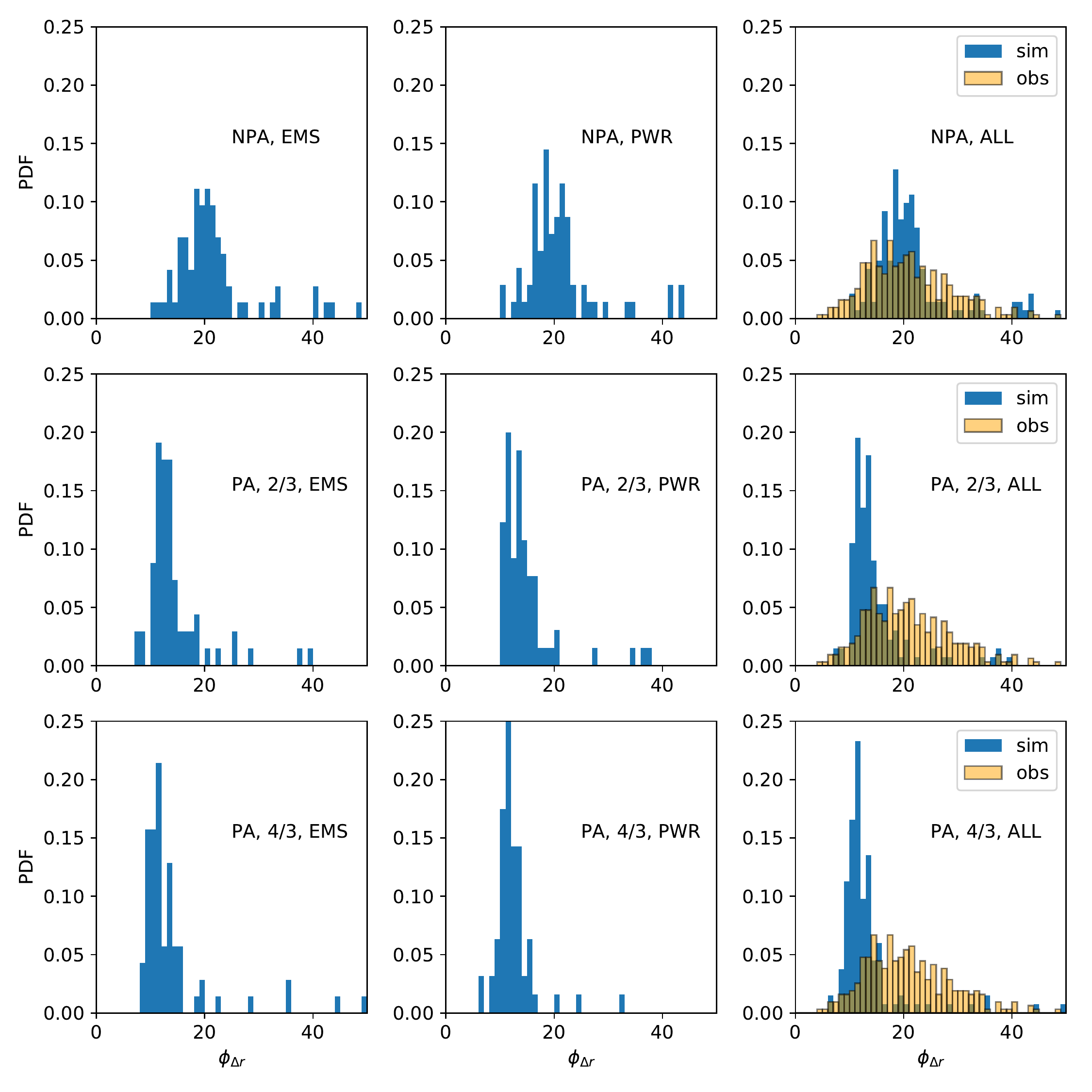}
	\caption{The probability distribution functions (PDFs) of the orbital spacing $\phi_{\Delta r}$ between the inner planet and the outer planet, where $\phi_{\Delta r} \equiv (a_{2} - a_{1}) / R_{H,1}$. Here $a_{2}$ and $a_{1}$ are the orbital semi-major axes of the outer planet and the inner planet, respectively, and $R_{H,1}$ is the Hill radius of the inner planet in the pair. The observational data (see text) are plotted in all the panels of the third column for comparison. The ordering of the subplot grid is the same as
Fig.~\ref{fig:mass_dist_hist_lin_x}.}
	\label{fig:rh_histogram}
\end{figure*}

To further investigate the orbital properties of innermost (planet 1) and next innermost (planet 2) planets, Figure~\ref{fig:rh_histogram} shows histograms of $\phi_{\Delta r,1} \equiv (a_{2} - a_{1}) / R_{H,1}$, i.e., the orbital separations of the planets normalised by the Hill radius of the
innermost planet. This metric has been considered
by \cite{2014ApJ...780...53C} and \cite{2016ApJ...816...19H} as a test of IOPF predictions, since the initial location of the second planet is expected to be at the displaced pressure maximum that is at least a few Hill radii away from the innermost planet and potentially much further away if DZIB retreat is controlled by X-ray ionizing photons penetrating from the host star and/or disk corona \citep{2016ApJ...816...19H}. The simulated planetary systems are seen to have typical values of $\phi_{\Delta r,1}\sim 10$ to 30, with the peak of the distribution systematically shifting from large values in the case of NPA to smaller values in the case of PA with $\beta=2/3$ and then the case of PA with $\beta=4/3$. This trend is explained simply as one driven by an increase in planetary masses when pebble accretion is included. More massive planets have larger Hill radii and so the distribution of $\phi_{\Delta r,1}$ shifts to smaller values.

Using the same exoplanet data as described above, in Figure~\ref{fig:rh_histogram} we also plot the observed distribution of $\phi_{\Delta r,1}$. For the masses of the observed innermost planets, which are needed for the estimate of $R_{H,1}$, we use an empirical density versus size relation
of the form:
\begin{equation}
    (\rho / {\rm g\:cm}^{-3}) = \begin{cases}
        10 & R_p \leq 1.4 R_{\oplus} \\
        17.8 (R_p/ 1.4 R_{\oplus})^{-1.86} & R_p > 1.4 R_{\oplus},
    \end{cases}
    \label{eq:rhoR}
\end{equation}
which has been calibrated from observed innermost, i.e., ``Vulcan'' planets by Brockett et al. (in prep.), building on methods of \cite{2015ApJ...798L..32C}.

The observed distribution of $\phi_{\Delta r,1}$ has values that extend from $\sim 5$ up to $\sim 40$, with a peak near about 15. While the position of the peak is similar to some of the simulated cases, i.e., with PA with either $\beta=2/3$ or 4/3, the simulated distributions are much narrower than the observed one. This, coupled with the results already seen for the distribution of $P_2/P_1$, indicate that the simulated planetary systems do not have orbital separation properties that match the observed STIPs.

Finally, we consider the property of the $M_{p}$ versus $a$ relation, i.e., how planet mass scales with orbital radius. The fiducial IOPF model predicts that each planet mass is set by shallow gap opening leading to truncation of pebble accretion with
\begin{equation}
M_p\propto a^{k_M}
\label{eq:Ma_relation}
\end{equation}
with $k_M=1/8$ \citep{2018ApJ...857...20H}. Note this is a slightly steeper relation than the value of $k_M=1/10$ that was first discussed by \cite{2014ApJ...780...53C}, with the difference arising from refined estimates of the gap opening mass \citep[see discussion in][]{2018ApJ...857...20H}. Nevertheless, such a scaling has the feature of being relatively shallow, so that planets within a given system have very similar masses that grow, on average, very slowly with semi-major axis. 

\begin{figure*}
	\centering
	\includegraphics[scale=0.8]{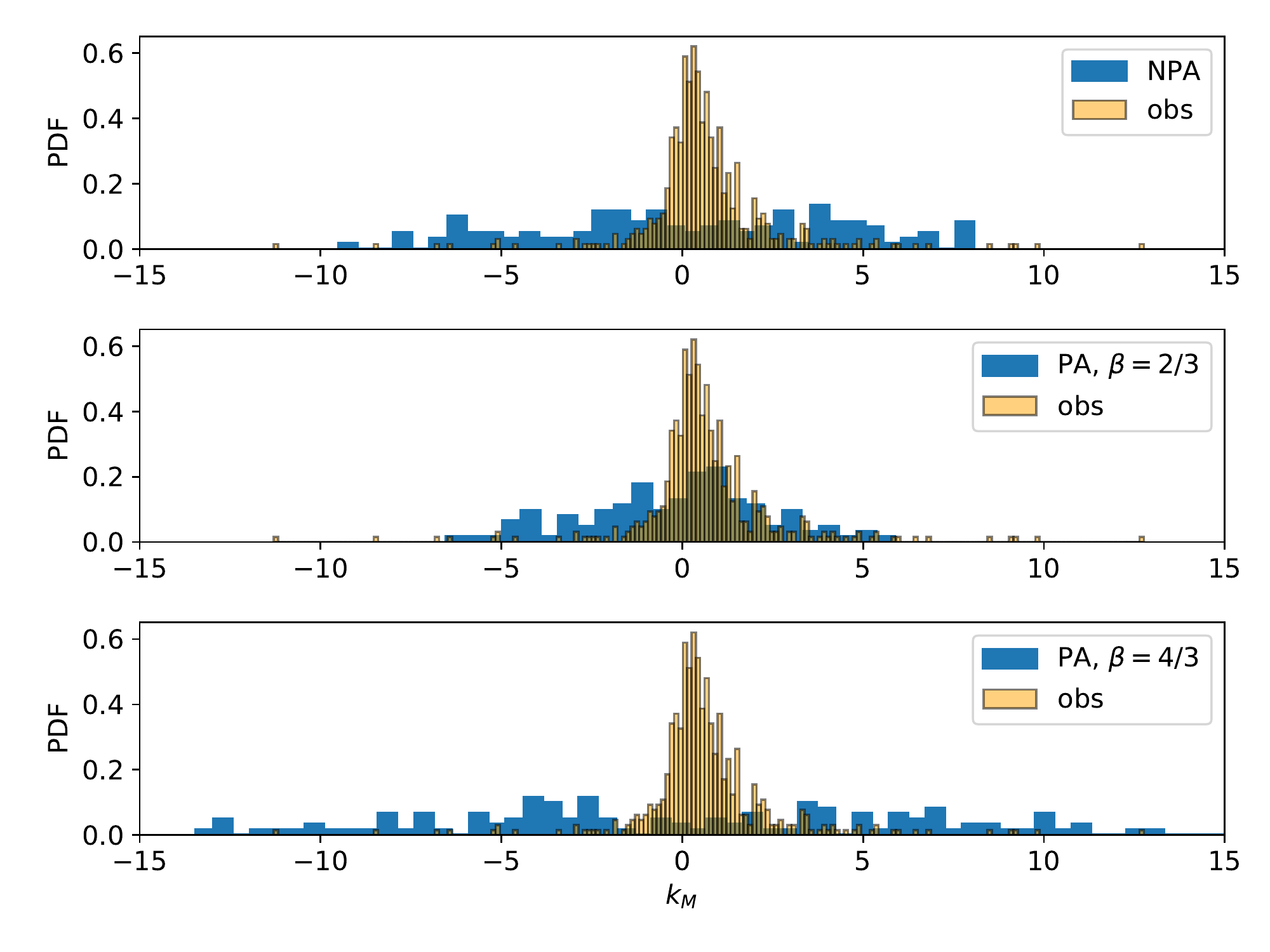}
	\caption{Normalized PDFs of the $k_M$ indices measured from the different models of the simulations. From top to bottom are shown: the models with no pebble accretion (NPA); the model with pebble accretion (PA) with $\beta=2/3$; and the models with PA with $\beta=4/3$. In each panel, the observed values of $k_M$ of STIPs inner planet pairs are shown in yellow (the observational data in each panel is identical). }
	\label{fig:km_histogram}
\end{figure*}

In Figure~\ref{fig:km_histogram} we show the distribution of $k_M$ values measured in our simulated planetary systems by fitting a power law of the form of eq.~\ref{eq:Ma_relation} to the properties of the innermost and next innermost planets. We find that generally $-10\lesssim k_M \lesssim 10$, i.e., there is a very wide range, including both positive and negative values. This reflects the fact that the primary and secondary have quite different masses, but are located at very similar average orbital distances from the host star, having formed from the same narrow planetesimal ring. In the models with pebble accretion with $\beta=4/3$ we see that the median $k_M\simeq-0.5$, indicating a preference for the primary to be the innermost planet. For the $\beta=2/3$ case, the median $k_M\simeq 0.2$, while in the NPA case it is close to zero.

We note that some of the most extreme values of $k_M$ may reflect the rare cases of tertiary planets being present as one of the innermost pair. The statistics of the $k_M$ indices are presented in Table~\ref{tab:km_stat}.

Figure~\ref{fig:km_histogram} also shows the observed values of $k_M$, based on the orbital radii and masses (via eq.~\ref{eq:rhoR}) of innermost planet pairs in the STIPs sample. 

We see that the observed values of $k_M$ have a median of $\simeq0.4$. We note there may be some systematic error in this result since it is based on using a density versus size relation that has been calibrated for Vulcan planets (Brockett et al., in prep.). The second innermost planets are likely to have systematically lower densities, e.g., due to greater degree of atmosphere retention, and so their masses may have been somewhat overestimated, which would lead to an overestimated value for $k_M$. Modulo this potential effect, we see that the mass versus orbital radius relation is relatively flat. As discussed above, a flat relation with $k_M=1/8$ is expected in the fiducial IOPF model in which planet mass is set by shallow gap opening leading to truncation of pebble accretion. Such $k_M$ values are also similar to the PA with $\beta=2/3$ case for planet pairs formed by oligarchic growth. However, as is clear from Figure~\ref{fig:km_histogram}, the observed distribution of $k_M$ values is much narrower than that seen in our oligarchic growth simulations, i.e., the observed dispersion is $\sigma(k_M)=1.3033$, which is much smaller than any of our simulated cases. Thus we conclude that, as with the $P_2/P_1$ and $\phi_{\Delta r,1}$ distributions, the distributions of $k_M$ indicate that the oligarchic growth model predictions are not consistent with those of inner planet pairs.

\begin{table}
	\caption{Statistics (mean, median and standard deviation) of planet mass versus semi-major axis $k_M$ indices derived from innermost and next innermost planet pairs in the simulation data (first three rows), compared with the STIPs observational data (fourth row). Note that in the observed STIPs, we discarded the 1\% most extreme values of $k_M$, so that the reported dispersion better reflects the core of the distribution. Such a truncation has minimal effect on the statistics of the simulation samples.}
	\centering
	\begin{tabular}{lrrr}
		\hline
		\hline
		Model & Mean $k_M$ & Median $k_M$ & $\sigma(k_M)$ \\
		\hline
		NPA         & 0.2046         & 0.0104        & 4.1880 \\
		\hline 
		PA, $\beta=2/3$ & -0.1624    & 0.1864        & 2.5207 \\
		\hline 
		PA, $\beta=4/3$ & -0.0046     & -0.4876        & 6.7553 \\
		\hline 
		Observed STIPs     & 0.6331     & 0.4351      & 1.3033 \\
		\hline 
	\end{tabular}
	\label{tab:km_stat}
\end{table}

\section{Discussion - Implications for IOPF}
\label{sec:discussion}

Inside-Out Planet Formation (IOPF) invokes a process of sequential planet formation from a series of pebble-rich rings that are located at the pressure trap of the dead zone inner boundary (DZIB). The very first planet to form in this model is the innermost, ``Vulcan'' planet with the DZIB located at the place in the disk where the midplane temperature reaches about 1,200~K allowing the thermal ionization of alkali metals \citep{2014ApJ...780...53C,2018ApJ...861..144M, 2021arXiv210212831J}. 
The fiducial model of IOPF {\it assumes} that the planet formation process from this pebble ring leads to creation of a single, dominant protoplanet, which then grows by continued pebble accretion until it reaches its pebble isolation mass, concurrent with shallow gap opening and DZIB retreat. Then the second planet forms by the same process from a second pebble ring.

The work of this paper has been to investigate whether the assumption of single dominant planet formation from the first pebble ring is valid in the case that this ring evolves to being a planetesimal ring, with a large part of the subsequent growth being controlled by planetesimal collisions. We have shown that while most of the mass does end up in the primary planet, a secondary planet always remains in the system with a mass of about 30\% to 65\% of that of the primary. First we discuss how robust this result is likely to be. Then we discuss the implications of the comparisons we have made with observed systems.

The results of our oligarchic growth simulations that lead to two planets, with the secondary being relatively massive, are insensitive to whether the initial planetesimals have an equal mass distribution or a power law mass distribution, to the initial width of the pebble ring and to whether pebble accretion is included or not. In this sense the result is relatively robust, although we have so far presented only a relatively limited exploration of the possible parameter space of initial planetesimal properties and pebble accretion rates. 

However, perhaps more importantly, certain features related to the physics of the problem have not yet been included. In particular, the protoplanets in our simulations do not feel any torques due to their interaction with the gas disk. Such torques have been included in some simulations of more radially extended disks of planetesimals \citep[e.g.,][]{2015A&A...578A..36O} and have been found to have significant effects on the dynamical evolution. Thus, it is possible that with the inclusion of such torques, the primary and secondary planets could, for example, eventually merge together. This possibility is made more likely by considering that the disk environment of the DZIB, which involves a steep drop in mass surface density as one moves inwards from the dead zone to the MRI-active zone, has been shown to be a global planetary migration trap in the case of single planet disk interactions \citep{2018ApJ...857...20H}. However, it remains to be shown that this planetary migration trap applies when two or more planets are orbiting and perturbing the disk in such a region.

Similarly, we have not included the effects of gas drag on the orbital motion of the planetesimals. One of the main effects that could be important is the damping of eccentricities, which in general would be expected to enhance the merger rate of the planetesimals and perhaps promote the merger of any final primary-secondary planetary pair. From the work of \cite{1976PThPh..56.1756A} we can evaluate the timescale of eccentricity damping, $t_{\rm e,damp}$, under conditions of the disk midplane near the DZIB relevant to IOPF via:
\begin{equation}
    t_{\rm e,damp} \equiv \frac{e}{|de/dt|} \simeq \left(\frac{8}{5}\right)^{1/2} \frac{t_0}{e},
\end{equation}
where the characteristic timescale $t_0$ is given by
\begin{equation}
    t_0 = \frac{1.427\times 10^4}{C_D} \left(\frac{M_p}{0.01 M_\oplus} \right)^{1/3} r_{\rm 0.1AU}^{-1} \rho_{-8}^{-1} t_{\rm orb},
\end{equation}
where $C_D$ is the dimensionless drag coefficient \citep[which for planetesimals greater in mass than $10^{18}\:$g is expected to be in the range 0.5 to 1.5,][]{1976PThPh..56.1756A}, $\rho_{-8} \equiv \rho / 10^{-8}\:{\rm g\:cm}^{-3}$ is the local gas density of the disk \citep[with this normalisation being to a typical value for the IOPF scenario;][]{2014ApJ...780...53C}
and $t_{\rm orb}$ is the local Keplerian orbital time. Thus we see that for $M_p=0.01\:M_\oplus$ and typical maximum values of $e\sim 0.1$, $t_{\rm e,damp} \sim 2 \times 10^5 t_{\rm orb} \simeq 6,000\:$yr and so gas drag would then have a significant effect in limiting the high-end of the eccentricity distribution of the planetesimal population. However, the effect becomes weaker for the more massive protoplanets, i.e., by a factor of about 5 for Earth-mass planets. The overall effects of both gas drag and migration torque effects on the final outcome of oligarchic growth from a planetesimal ring are important topics for future investigation.


A second assumption in our modeling is that in the simulations with pebble accretion there is no dependence on orbital radius of the pebble accretion rate. For example, one may expect that in certain circumstances the outermost planet may be able to intercept most of the pebble flux and prevent pebbles from reaching inner planets. However, such efficient interception is only expected for relatively massive protoplanets. Still if this regime was to be reached, then it could lead to runaway growth of a single planet that would reduce the importance of any remnant secondary planet.

A third assumption of our modeling that may have an impact on the result is that the planetesimal population is all in place from the outset at $t=0$. In reality, the planetesimals would form gradually from the pebble ring and the earliest object to form would then potentially have a significant head start in the growth process that might affect the entire subsequent evolution. While our models with a power law distribution of initial planetesimal masses go some way to addressing this concern, the question requires a more thorough investigation with simulations that involve gradual introduction of planetesimals and a study of the effects of different rates of their formation.

The comparison of the properties of the primary and secondary planets formed in our simulations with those of the innermost pairs of planets in real systems shows several discrepancies, i.e.,  the observed planets have a broader distribution of period ratios ($P_2/P_1$), a broader distribution of orbital separations normalized by inner planet Hill radius ($\phi_{\Delta r, 1}$) and a narrower range of planet mass versus semi-major axis scaling indices ($k_M$). These results indicate that the observed planets in STIPs did not form via a process of oligarchic growth from a narrow planetesimal ring---at least not as we have modelled here. Thus, if IOPF is to be a valid theory for the formation of STIPs, then it still remains to be demonstrated that it can avoid such an oligarchic growth phase or that if such a phase is involved that it does not lead to remnant secondary planets with properties that have been seen in the simulations presented here.

\section{Conclusions}
\label{sec:conclusions}

The Inside-Out Planet Formation (IOPF) theory predicts that Earth to Super-Earth mass planets form {\it in situ} from a sequential series of pebble-rich rings that are coincident with the pressure trap of the dead zone inner boundary with an MRI-active inner zone.

IOPF assumes a single dominant planet emerges from such pebble rings, with the final mass then set by self-truncation of pebble accretion once the planet becomes massive enough to open a shallow gap, displacing the local pressure maximum and initiating retreat of the DZIB to larger radii. However, it has not yet been investigated exactly how planet formation from the pebble ring may actually occur.

In this paper, using direct $N$-body simulations and assuming that the drag and torques due to interaction with the gas disk are negligible, we have studied the collisional evolution of planetesimal rings that may have formed from the first, i.e., innermost, IOPF pebble ring. Our main conclusions are summarized as follows:

\begin{enumerate}

\item All planetesimal rings that we have investigated undergo oligarchic growth. Regardless of the initial conditions, such as the initial planetesimal mass function and initial width of the planetesimal ring, and regardless of whether pebble accretion is also included, the collisional evolution of planetesimals generally results in 2 (and occasionally 3) oligarchs after 1~Myr of evolution. These oligarchs are settled in nearly coplanar and circular orbits. The most massive planetesimal typically consists of 2/3 of the total initial mass, i.e., with the secondary having 30\% to 65\% of that of the primary.

\item Independent of the initial conditions and pebble accretion assumptions, only a small fraction of oligarchs are captured in orbits with low-order mean-motion resonance, with 9/7 being the most important. Most adjacent oligarchs have orbital period ratio of $1.1-1.4$.

\item Independent of the initial conditions, the mass growth of oligarchs is highly efficient, with most planetesimals having undergone collision within $\sim10^5\:$yr.

\item The observed properties of innermost planet pairs in STIPs are inconsistent with the properties of our simulated planetary systems, i.e.,  the observed planets have a broader distribution of period ratios ($P_2/P_1$), a broader distribution of orbital separations normalized by inner planet Hill radius ($\phi_{\Delta r, 1}$) and a narrower range of planet mass versus semi-major axis scaling indices ($k_M$). This indicates that the simulated oligarchic growth phase leading to survival of a relatively massive secondary cannot explain the observed innermost planets of STIPs.

\item For IOPF to be a valid theory to explain the observed STIPs, further work is needed to investigate whether a single dominant planet, i.e., with insignificant secondary, can emerge from the pebble ring. Improved modeling that includes a treatment of planet-disk interactions,  i.e., migration torques and gas drag, and gradual formation of the planetesimals from the pebble ring are among the next aspects of this problem to be investigated.

\end{enumerate}

\section*{Acknowledgements}
We thank the anonymous referee for his/her comments that helped to improve the manuscript considerably. We acknowledge useful discussions with Genesis Brockett, Sourav Chatterjee, Willy Kley, Eiichiro Kokubo, Chris Ormel and Jon Ramsey. 
This work was partially supported by SURF (Dutch National Supercomputing Center), and partially supported by the Netherlands Research Council NWO (grants \#643.200.503, \#639.073.803 and \#614.061.608) by the Netherlands Research School for Astronomy (NOVA). JCT acknowledges support from NASA ATP grant 80NSSC19K0010.

\section*{Data Availability}
The simulation data underlying this article will be shared on reasonable request to the corresponding author. The exoplanet observational data is publicly available and can be downloaded directly from \href{http://exoplanet.eu/}{exoplanet.eu}.




\bibliographystyle{mnras}
\bibliography{refs} 



\appendix

\section{Gravitational focusing enhanced pebble accretion in thick pebble disks}
\label{appendix:thick_disk}

Since motions of pebbles are coupled with those of the gas, the
scaleheight of the pebble layer can be stirred up by turbulence in the
disk. Following \cite{2007astro.ph..1485A}, the thickness of a pebble
layer, $z$, due to turbulence is given implicitly via
\begin{equation}
	\frac{h^2}{z^2}e^{-z^2/h^2} = \frac{\pi \rho_{d} a}{2 \alpha \Sigma},
	\label{eq:thickness}
\end{equation}
where $h$ is the vertical scaleheight of the disk, $\rho_{d}$ is the
mean density of pebbles, $\alpha$ is the Shakura-Sunyaev viscosity
parameter of the disk, and $\Sigma$ is the mass surface density of the
disk. 
As we will see, we are typically in a regime where the pebble scaleheight is small compared to that of the gas so that $e^{-z^2/h^2} \simeq 1$ and
\begin{equation}
    \frac{z}{h} \simeq \left(\frac{2\alpha \Sigma}{\pi \rho_d a}\right)^{1/2}= 0.146 \alpha_{-4}^{1/2} \Sigma_3^{1/2} \rho_{d,3}^{-1/2} a_1^{-1/2},
    \label{eq:thickness2}
\end{equation}
where $\alpha_{-4}\equiv \alpha/10^{-4}$, $\Sigma_3\equiv \Sigma/10^3\:{\rm g\:cm^{-2}}$, $\rho_{d,3}\equiv \rho_d/3\:{\rm g\:cm}^{-2}$ and $a_1=a/1\:{\rm cm}$.

In the IOPF disk model of \cite{2018ApJ...857...20H} (see their
Fig. 1), the aspect ratios of the fiducial disk models with gas
accretion rates of $\dot{m}=10^{-9}\:M_\odot\:{\rm yr}^{-1}$ at
$r=0.1$~au are $h/r \simeq 3.3 \times 10^{-2}$ and $h/r \simeq 2.2
\times 10^{-2}$ for values of DZIB region $\alpha = 10^{-4}$ and
$\alpha = 10^{-3}$, respectively. The mass surface densities at
$r=0.1$~au are $\Sigma \simeq 4 \times 10^3$~$\mathrm{g\:cm}^{-2}$ and
$\Sigma \simeq 10^3$~$\mathrm{g\:cm}^{-2}$ for $\alpha = 10^{-4}$ and
$\alpha = 10^{-3}$, respectively. Thus, for typical pebble radii of $a= 0.1,
1, 10$~cm, from eq.~\ref{eq:thickness2} we obtain the thickness of the
pebble layer due to turbulent stirring as
\begin{equation}
	z = 
	\begin{cases}
		(30, 9.6, 3.0) \times 10^{-4} ~\mathrm{au} & \alpha=10^{-4} \\
		(32, 10, 3.2) \times 10^{-4} ~\mathrm{au} & \alpha=10^{-3}
	\end{cases},
\end{equation}
respectively, which far exceed the radius of a planetesimal, i.e.,
a 1 Earth mass planestimal in our modeling has a radius of 8900~km ($=6.0\times10^{-5}\:$au).
As such, we consider the pebble layer at the DZIB as being
relatively thick compared to planetesimal size, and that pebble
accretion is then permitted from arbitrary directions.

Pebble accretion may be enhanced by gravitational focusing. Consider a
pebble, with a mass that is negligible compared to that of the
planetesimal of mass $M$ and radius $R$, approaching from an arbitrary
direction, ignoring the background potentials from the host star and
the protoplanetary disk, gas drag, and perturbations from other
pebbles.  Then energy conservation during the interaction implies
\begin{equation}
	v_{\infty}^2/2 = v^2/2 - GM/R,
\label{eq:energy_consv}
\end{equation}
where
$v_{\infty}$ is the relative velocity at infinity, $v$ is the speed
when the two particles have a distance $R$, i.e., at the point of
collision.
In the limit of conservation of angular momentum for the
planetesimal-pebble interaction,
\begin{equation}
	 b v_{\infty} =   R v,
\label{eq:am_consv}
\end{equation}
where $b$ is the impact parameter. Combining
eqs.~(\ref{eq:energy_consv}) and (\ref{eq:am_consv}) yields
\begin{equation}
	v_{\infty}^2/2 +  G M/R =  b^2 v_{\infty}^2/(2R^2).
\end{equation}
Expressing in terms of the escape speed from the planetesimal, $v_e =
(2GM/R)^{1/2}$, we have
\begin{equation}
	b^2 = R^2 \left( 1 + \frac{v_{e}^2}{v_{\infty}^2} \right).
\end{equation}
Thus the gravitational-focusing-enhanced cross section is
\begin{equation}
	\Gamma = \pi b^2 = \pi R^2 \left( 1 + \frac{v_{e}^2}{v_{\infty}^2} \right)
\end{equation}
In the limit $v_{e}^2/v_{\infty}^2 \gg 1$, then
\begin{equation}
	\Gamma \simeq \pi R^2 \frac{2GM}{R v_{\infty}^2} = \frac{2 \pi RGM}{v_{\infty}^2} \propto M^{4/3},
\end{equation}
assuming planetesimal density and $v_\infty$ are independent of $M$.

Therefore, the individual pebble accretion rate for the thick pebble disk case is
(cf. Eq.23 of \cite{1993prpl.conf.1061L})
\begin{equation}
	\dot{M}_i = \rho \sigma \Gamma(M_i) = k M_i^{4/3}.
\end{equation}
Similarly, for the thin pebble disk case, we have $\Gamma\propto R v_e^2\propto M$.

However, it is possible that in a Keplerian protoplanetary disk
environment that factors such as $v_{\infty}$ also have a mass
dependence, e.g., if set by the disk shear velocity at the tidal
radius of the protoplanet. Furthermore, the assumptions of energy and
angular momentum conservation may not be accurate if gas drag forces
are significant during the infall phase. Numerical simulations that
treat these effects have been carried out to investigate pebble
accretion.

As a result of these uncertainties, in our study we explore three
different cases: no pebble accretion; pebble accretion with
$\Gamma\propto M^{2/3}$; and $\Gamma\propto M^{4/3}$.



\bsp	
\label{lastpage}
\end{document}